\documentclass[letters,a4paper,fleqn,usenatbib,useAMS]{mnras}

\usepackage{psfrag,color,epsfig}

\usepackage{graphicx}	
\usepackage{amsmath}	
\usepackage{amssymb}	
\usepackage{multicol}	
\usepackage{bm}		
\usepackage{pdflscape}	
\usepackage{multirow}
\usepackage{tabularx}  
\usepackage{caption}

\usepackage{physics}
\usepackage{xspace}
\usepackage{soul} 
\usepackage[T1]{fontenc}
\usepackage{ae,aecompl}

\usepackage{newtxmath}

\def\HI{\ion{H}{I}~}
\def\Hi{\ion{H}{I}}

\def\xHI{x_{\rm \ion{H}{I}}}
\def\xb{\bar{x}_{\rm \ion{H}{I}}}
\def\Tb{T_{\rm b}}

\def\k{{\bm{k}}}

\def\x{\bm{x}}
\def\r{\bm{r}}

\def\thetavec{\bm{\theta}}
\def\cl{{\mathcal C}_{\ell}}
\def\c{{\mathcal C}}
\def\n{\hat{\bm{n}}}
\def\v{\bm{v}}

\title[Reionization history]{A method to determine the evolution history of the mean neutral Hydrogen fraction}

\author[R. Mondal et al.]{Rajesh Mondal$^1$\thanks{E-mail: \href{mailto:Rajesh.Mondal@sussex.ac.uk}{Rajesh.Mondal@sussex.ac.uk}},
Somnath Bharadwaj$^2$, Ilian T. Iliev$^1$, Kanan K. Datta$^3$,
\newauthor Suman Majumdar$^{4, 5}$, Abinash K. Shaw$^2$ and Anjan K. Sarkar$^2$\\
$^1$ Astronomy Centre, Department of Physics and Astronomy, University of 
Sussex, Brighton BN19QH, UK\\
$^2$ Department of Physics \& Centre for Theoretical Studies, Indian 
Institute of Technology Kharagpur, Kharagpur 721302, India\\
$^3$ Department of Physics, Presidency University, 86/1 College Street, 
Kolkata 700073, India\\
$^4$ Centre of Astronomy, Indian Institute of Technology Indore, 
Simrol, Indore 453552, India\\
$^5$ Department of Physics, Blackett Laboratory, Imperial College, London SW7 2AZ, UK}

\date{\today}
\pubyear{2018}

\begin{document}
\label{firstpage}
\pagerange{\pageref{firstpage}--\pageref{lastpage}}
\maketitle


\begin{abstract}
The light-cone (LC) effect imprints the cosmological evolution of the 
redshifted 21-cm signal $T_{\rm b} (\n, \nu)$ along the frequency axis 
which is the line of sight~(LoS) direction of an observer. The effect 
is particularly pronounced during the Epoch of Reionization~(EoR) when 
the mean hydrogen neutral fraction $\xb(\nu)$ falls rapidly as the 
universe evolves. The multi-frequency angular power spectrum~(MAPS) 
$\cl(\nu_1,\nu_2)$ quantifies the entire second-order statistics of 
$T_{\rm b}(\n,\nu)$ considering both the systematic variation along 
$\nu$ due to the cosmological evolution and also the statistically 
homogeneous and isotropic fluctuations along all the three spatial 
directions encoded in $\n$ and $\nu$. Here we propose a simple model 
where the systematic frequency $(\nu_1,\nu_2)$ dependence of 
$\cl(\nu_1,\nu_2)$ arises entirely due to the evolution of $\xb(\nu)$. 
This provides a new method to observationally determine the 
reionization history. Considering a LC simulation of the EoR 21-cm 
signal, we use the diagonal elements $\nu_1=\nu_2$ of $\cl(\nu_1,\nu_2)$
to validate our model. We demonstrate that it is possible to recover
the reionization history across the entire observational bandwidth
provided we have the value $\xb$ at a single frequency as an external
input. 
\end{abstract}




\section{Introduction}
\label{sec:intro}
Observations of the redshifted 21-cm signal from neutral hydrogen~(\Hi) are 
the most promising probe of the epoch of reionization~(EoR). A considerable 
amount of effort is underway to detect the EoR 21-cm signal using ongoing and 
upcoming radio interferometric experiments e.g. GMRT \citep{paciga13}, LOFAR 
\citep{haarlem13,yatawatta13}, MWA \citep{bowman13,tingay13,dillon14}, PAPER 
\citep{parsons14,ali15,jacobs14}, SKA \citep{mellema13,koopmans15} and HERA 
\citep{deboer17}.

Using the redshifted \HI 21-cm signal one can, in principle, map the \HI
distribution in the intergalactic medium~(IGM) in 3D with the line of
sight~(LoS) axis being the frequency (or redshift). However, an
observer's view of the universe is restricted to the backward light-cone,
and the \HI 21-cm signal evolves along the line of sight~(LoS). This
gives rise to the `light-cone'~(LC) effect which has a significant impact
on the EoR 21-cm signal and its various statistics. This has been taken 
into account by \citet{barkana06} and \citet{zawada14} while modelling 
the LC anisotropies in the two-point correlation function. 
\citet{datta12,datta14} and \citet{la-plante14} have examined the impact 
of this effect on the EoR 21-cm 3D power spectrum which is the primary 
observable of the first generation of radio interferometers.

Another important line of sight~(LoS) effect is the redshift space
distortion~(RSD) due to the peculiar velocities of \Hi. Similar to 
the LC effect RSD introduces anisotropies in the 21-cm signal 
\citep{bharadwaj04} along the LoS. Although, there has been 
substantial effort invested in including the RSD in EoR simulations 
by \citet{mao12,majumdar13,majumdar16,jensen13}, the problem of how 
to properly include the peculiar velocities of \HI in LC simulation 
was addressed by \cite{mondal18}.

The statistical homogeneity (or ergodicity) along the LoS gets destroyed 
by the LC effect. One of the main problems regarding the interpretation 
of the EoR 21-cm signal through its 3D power spectrum $P(\k)$ lies in 
the signal's non-ergodic nature. The 3D power spectrum $P(\k)$ assumes 
that the signal is ergodic and periodic, thus it provides a biased 
estimate of the statistics of EoR signal \citep{trott16}. In contrast, 
the multi-frequency angular power spectrum (hereafter MAPS) 
$\cl(\nu_1,\nu_2)$ \citep{datta07a} does not have any such intrinsic 
assumption in its definition. \citet{mondal18} have demonstrated that 
the entire second-order statistics of the non-ergodic LC EoR signal can 
be expressed by the MAPS.

In this Letter, we demonstrate, as a proof of concept, how one can use the 
intrinsic non-ergodicity of the light-cone EoR 21-cm signal  to uncover 
the underlying reionization history. The reionization history is one of 
the most sought-after outcomes of any experiment aiming to observe the EoR. 
We propose and validate a formalism   whereby the measured MAPS can be 
used to extract the reionization history in a model independent manner. 
In this Letter, we have used the Planck$+$WP best fit values of 
cosmological parameters \citep{planck14}.



\section{Simulating the light-cone 21-cm signal from the E\lowercase{o}R}
\label{sec:LCsim}
In this section, we briefly summarise the simulation technique used for
generating the light-cone EoR 21-cm signal. The reader is referred to the
Section~2 of \citet{mondal18} for a detailed description of the simulations.
Here, we have considered a region that spans the comoving distance range 
$r_{\rm n}=9001.45~{\rm Mpc}$ (nearest) to $r_{\rm f}=9301.61~{\rm Mpc}$ 
(farthest), which correspond to the frequencies $\nu_{\rm n}=166.91~{\rm Mhz}$
and $\nu_{\rm f}=149.04~{\rm Mhz}$, respectively. We have simulated snapshots
of the \HI distribution ({\it coeval} cubes) at 25 different co-moving
distances $r_{\rm i}$ in the aforesaid $r$ range (see Fig.~2 of 
\citealt{mondal18}) which were chosen so that the mean neutral Hydrogen 
fraction $\xb$ varies approximately by an equal amount in each interval. 

We have used semi-numerical simulations to generate the coeval ionization
cubes with comoving volume $V=[300.16~{\rm Mpc}]^3$. These simulations 
involve three main steps. First step involves a particle-mesh~(PM) 
$N$-body code to simulate the dark matter distribution. The $N$-body 
run has $4288^3$ grids with $0.07\,{\rm Mpc}$ grid spacing using 
$2144^3$ dark matter particles (particle mass $1.09 \times 10^8\,M_{\sun}$). 
In the next step, a Friends-of-Friends~(FoF) algorithm is used to 
identify collapsed halos in the dark matter distribution. A fixed 
linking length of $0.2$ times the mean inter-particle distance is 
used for the FoF and we have set the criterion that a halo should 
have at least $10$ dark matter particles. In the third and last step, 
an ionization field following an excursion set formalism 
\citep{furlanetto04a} is produced. For this we have adopted the 
ionization parameters $\{N_{\rm ion},\,M_{\rm halo, min}, 
\,R_{\rm mfp}\}=\{23.21,\, 1.09\times 10^9~M_{\sun},\, 20~{\rm Mpc}\}$, 
identical to \citet{mondal17}. This final step  closely follow the 
assumption of homogeneous recombination adopted by \citet{choudhury09b}. 
The \HI distribution in our simulations are represented by particles 
whose \HI masses were calculated from the neutral Hydrogen fraction 
$\xHI$ interpolated from its eight adjacent grid points. The positions, 
peculiar velocities and \HI masses of these particles are then saved 
for each such coeval cube. 

To construct the LC map, we slice the coeval maps at $25$ different 
radial distance $r_{\rm i}$, and construct the LC map for the region 
between $r_{\rm i}$ to $r_{\rm i+1}$ with the \HI particles from 
corresponding slices of the coeval snapshot. Finally, we map the \HI 
particles within the LC box from $\r= r \n$ to observing frequency 
$\nu$ and direction $\n$ which are the appropriate variables for the
observations of redshifted 21-cm brightness temperature fluctuations
$\delta \Tb (\n,\,\nu)$ in 3D. Note that for this mapping, the
cosmological expansion and the radial component of the \HI peculiar
velocity $\n\cdot\v$ together determine the observed frequency $\nu$ 
for the 21-cm signal originating from the point $\n r$. Our LC box is 
centered at the co-moving distance $r_{\rm c} = 9151.53~{\rm Mpc}$ 
($\nu_{\rm c} = 157.78 \, {\rm MHz}$) which correspond to the redshift 
$\approx 8$. The mass-averaged \HI fraction $\xb$ at the centre of 
the LC simulation is $\approx 0.51$, and it changes from 
$\xb \approx 0.65$ (at farthest end) to $\xb \approx 0.35$ (at 
nearest end), following the reionization history Fig.~2 of 
\citet{mondal18}.


\section{Modelling the multi-frequency angular power spectrum}
\label{modeling}
The issue under consideration here is `How to quantify the statistics of 
the non-ergodic EoR 21-cm signal $\delta \Tb (\n,\,\nu)$ in 3D?'. We know 
that the LC effect makes the cosmological 21-cm signal ($\delta T_{\rm b} 
(\n,\,\nu)$) evolve significantly along the LoS direction $\nu$. The 3D 
power spectrum $P(\k)$ is not accurate when the statistical properties of 
the signal evolve along a specific direction. Additionally, the Fourier 
transform imposes periodicity on the signal, an assumption that cannot be 
justified along the LoS when the LC effect has been taken into account. 
As a consequence, the 3D power spectrum fails to quantify the entire 
information in the signal and gives a biased estimate of the statistics 
\citep{trott16, mondal18}. In contrast to this the MAPS $\cl(\nu_1,\nu_2)$ 
quantifies the entire second order statistics of the EoR 21-cm signal 
even in the presence of the LC effect \citep{mondal18}.

The redshifted 21-cm brightness temperature fluctuations are decomposed 
into spherical harmonics as
\begin{equation}
\delta T_{\rm b} (\n,\,\nu)=\sum_{\ell,m} a_{\ell {\rm m}} (\nu) \,
Y_{\ell}^{\rm m}(\n)\, ,
\label{eq:a1}
\end{equation}
and these are used to define the MAPS \citep{datta07a} using 
\begin{equation}
\cl(\nu_1, \nu_2) = \big\langle a_{\ell {\rm m}} (\nu_1)\, a^*_{\ell 
{\rm m}} (\nu_2) \big\rangle\, .
\label{eq:cl}
\end{equation}
This takes into account the assumption that the EoR 21-cm signal is
statistically homogeneous and isotropic with respect to different
directions in the sky, however it does not assume the signal to be 
statistically homogeneous along the LoS direction $\nu$. Considering 
the particular situation where the signal is ergodic (statistically 
homogeneous) along the LoS, we have $\cl(\nu_1, \nu_2)=\cl(\Delta \nu)$ 
{\it i.e.} it depends only on the frequency separation 
$\Delta \nu=\mid \nu_1-\nu_2 \mid$.

Under the assumption that the \HI spin temperature is much larger 
than the CMB temperature i.e. $T_{\rm s} \gg T_{\gamma}$, the 
redshifted 21-cm brightness temperature fluctuations  (eq.~4 and A5 
of \citealt{bharadwaj05}) can be expressed as \citep{mondal18}
\begin{equation}
T_{\rm b} (\n, \nu) = \bar{T}_0 \, \frac{\rho_{\Hi}}{\bar{\rho}_{\rm H}}\,
\left( \frac{H_{0} \nu_{\rm e}}{c} \right)  \, \bigg| \frac{\partial
  r}{\partial \nu} \bigg| \,, 
\label{eq:bt}
\end{equation}
where
\begin{equation}
\bar{T}_0 = 4.0 {\rm mK}\, \left( \frac{\Omega_{\rm b} h^2}{0.02}\right)
\left( \frac{0.7}{h} \right) \,,
\end{equation}
${\rho_{\Hi}}/{\bar{\rho}_{\rm H}}$ is the ratio of the neutral hydrogen 
density to the mean hydrogen density, and $r$ refers to the  comoving 
distance from which the redshifted \HI emission, observed at frequency 
$\nu$, is originated. The factors ${\rho_{\Hi}}/{\bar{\rho}_{\rm H}}$ and 
${\partial r}/{\partial \nu} $ both evolve along the LoS direction 
($\nu$ or $z$) due to a variety of factors including the  evolution 
of various quantities pertaining to the background cosmological model 
and the growth of density perturbation in $\rho_{\Hi}$. However during 
the EoR the evolution of the mean mass weighted neutral hydrogen 
fraction $\xb={\bar{\rho}_{\Hi}}/{\bar{\rho}_{\rm H}}$ by far dominates 
over the other factors that  cause $T_{\rm b} (\n, \nu)$ to evolve 
along the LoS direction.  Based on this we propose a model 
\begin{equation}
\cl(\nu_1,\nu_2)=\xb(\nu_1) \, \xb(\nu_2) \, \cl^{{\rm E}}(\Delta \nu)\,, 
\label{eq:a4}
\end{equation}
where $\cl^{{\rm E}}(\Delta \nu)$ is ergodic along the LoS, and the 
factor $\xb(\nu_1) \, \xb(\nu_2)$ which accounts for the evolution of 
the mean hydrogen neutral fraction breaks the ergodicity along the LoS.
We expect the above relation to hold at small scales where the HI density 
traces the underlying DM density. However, at scales larger than the 
typical bubble size the evolution is expected to be dominated by the 
evolution of bubble sizes and the above relation may not stay valid in 
that regime. This will provide a handle to measure the evolution of the 
\HI neutral fraction  $\xb(z)$ as reionization proceeds. Unfortunately 
this will only allow us to determine the ratio $\xb(z_2)/\xb(z_1)$ at 
two different epochs, and it will not allow us to uniquely determine 
$\xb(z_1)$ or $\xb(z_2)$. For the purpose of this Letter we consider 
\begin{equation}
\frac{\xb(z_2)}{\xb(z_1)}=\sqrt{\frac{\cl(\nu_2,\nu_2)}{\cl(\nu_1,\nu_1)}} \, ,
\label{eq:a5}
\end{equation}
which does not uniquely determine the \HI reionization history. However 
the reionization history is uniquely specified if we combine these 
measurements with a single  measurement of $\xb$ at any particular 
epoch say $z_1$ using an independent method (e.g. \citealt{majumdar12}).


\begin{figure*}
\centering
\includegraphics[width=1.01\textwidth, angle=0]{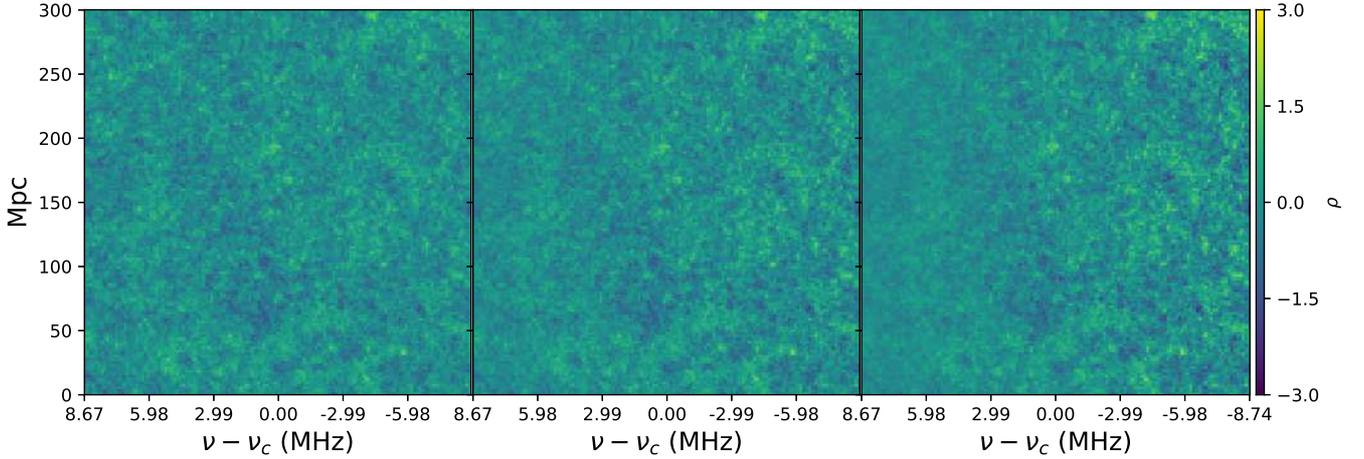}
\caption{This shows a section through one realization of the different 
Gaussian brightness temperature fluctuations fields for $a=0$ (ergodic), 
$0.5$ and $1.0$ (see eq.~\ref{eq:func}). In the panels, the value of $a$ 
increases from the left to the right.}
\label{fig:maps}
\end{figure*}

\section{Validating our model}
\label{test}
As a first step towards validating our model we consider a situation where 
the brightness temperature fluctuations are, by construction, of the form 
\begin{equation}
\delta T_{\rm b} (\n, \nu)=f(\nu) \times \delta_{\rm e}(\n,\nu)\,,
\end{equation}
where $f(\nu)$ is a known function and $\delta_{\rm e}(\n,\nu)$ is a random 
field which is isotropic in $\n$ and ergodic in $\nu$. Using this we investigate 
whether our method of analysis can determine $f(\nu)$ from the estimated 
$\cl(\nu_1,\nu_2)$. Here we have simulated $10000$ statistically independent 
realizations of homogeneous and isotropic Gaussian random fields 
$\delta_{\rm e}(\x)$ corresponding to a realisation of the CDM power 
spectrum $P_{\rm DM}(k)$. Working in the regime where the flat sky 
approximation holds true, we have converted the comoving displacement 
with respect to the centre of the box $\x=(\x_{\perp},x_{\parallel})$ 
to angle and frequency respectively using $\thetavec=\x_{\perp}/r$ and 
$\nu-\nu_{\rm c}=x_{\parallel}/\r^{\prime}$ where we assume that the 
centre of the simulation box is located at a redshift 
$z_{\rm c}=1420 \, {\rm MHz}/\nu_{\rm c}$ with corresponding comoving 
distance $r$ and with $r^{\prime}= {\rm d}r/{\rm d}\nu$ evaluated at 
$\nu_{\rm c}$. The resulting $\delta_{\rm e}(\thetavec,\nu)$ is 
statistically isotropic in $\thetavec$ and ergodic in $\nu$. The 
ergodicity along the LoS is broken by the function $f(\nu)$ which 
we have assumed to be of the form
\begin{equation}
f(\nu)=1-a\left( \frac{\nu-\nu_{\rm c}}{B} \right)\,.
\label{eq:func}
\end{equation}
Here $f(\nu)$ is a linear function which has value $f(\nu_{\rm c})=1$ at the 
centre of the frequency bandwidth $B$, and it has values $f=1-a/2$ and 
$f=1+a/2$ at the nearest and furthest edges of the band. In principle, one 
can choose different forms of $f(\nu)$. The aim here 
is to mimic a situation where we are analysing observations of a part of 
the reionization history where the evolution of the neutral fraction is 
approximately linear (see Fig.~2 of \citealt{mondal18}). Different values 
of $a$ correspond to different values of the slope or equivalently 
different values of the reionization rate. The different panels of 
Fig.~\ref{fig:maps} show $\delta T_{\rm b}(\thetavec,\nu)$ for a single 
realization of $\delta_{\rm e}(\x)$ considering different values of $a$.  

\begin{figure}
\psfrag{ratio}[c][c][1][0]{\large $A\,\sqrt{\cl(\nu)/\bar{\cl}}$}
\psfrag{nu}[c][c][1][0]{\large $\nu - \nu_{\rm c}$\,~MHz}
\psfrag{coeval}[c][c][1][0]{\large Ergodic~}
\psfrag{a=0.1}[c][c][1][0]{\large $a=0.1$~}
\psfrag{a=0.2}[c][c][1][0]{\large $a=0.2$~}
\psfrag{a=0.5}[c][c][1][0]{\large $a=0.5$~}
\psfrag{a=1.0}[c][c][1][0]{\large $a=1.0$~}
\centering
\includegraphics[width=0.49\textwidth, angle=0]{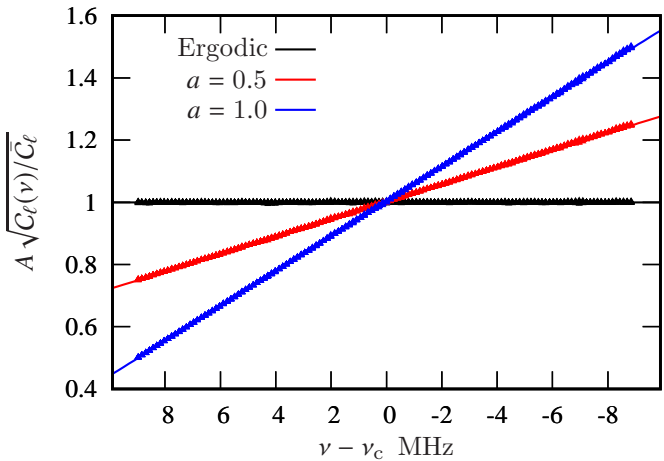}
\caption{This shows $A\,\sqrt{\cl(\nu)/\bar{\cl}}$ for different values of 
$a$ as shown in the figure. The points show results from our simulations 
and the solid lines show the function $f(\nu)$~(eq.~\ref{eq:func}).}
\label{fig:cl_ratio}
\end{figure}

We have used the simulated $\delta T_{\rm b}(\thetavec,\nu)$ to estimate 
$\cl(\nu_1,\nu_2)$ in the flat sky approximation \citep{mondal18}. 
Here, we focus on the diagonal elements $\nu_1 = \nu_2$ where the
MAPS signal peaks. In principle, one can use the full information 
contained in MAPS matrix $\cl(\nu_1, \nu_2)$ to analyse the results. 
However for simplicity we have only considered the diagonal terms.
We have used the ratio $A \sqrt{\cl(\nu)/\bar{\cl}}$ to determine 
$f(\nu)$ from our simulations. Here $\cl(\nu) \equiv \cl(\nu,\nu)$, 
$\bar{\cl}=B^{-1} \, \int^{B/2}_{-B/2} \, \cl(\nu) \, d \nu$ and $A$ 
is a normalisation constant whose value has to be externally specified. 
Here we use the prior information that $f(\nu_{\rm c})=1$ to decide the value 
of $A$. Fig.~\ref{fig:cl_ratio} shows the ratio $A \sqrt{\cl(\nu)/
\bar{\cl}}$ evaluated at different $\ell$ values which all have 
been shown as function of $\nu - \nu_{\rm c}$. We have used $10$ 
equally spaced logarithmic $\ell$ bins. We find that the ratio is 
independent of $\ell$ i.e. they all overlap. We also see that the 
ratio is able to correctly recover the functional form $f(\nu)$ 
from the simulations shown in Fig.~\ref{fig:maps}. This validates 
our method of analysis. 


We next apply the same method to our LC simulations to test if our model 
(eq.~\ref{eq:a4}) actually holds for the simulated EoR 21-cm signal.
The light-cone EoR 21-cm signal is undoubtedly non-ergodic along the LoS. 
An earlier work (Fig.~9 of \citealt{mondal18}) demonstrates that 
$\cl(\nu)/\bar{\cl} - 1$ shows a systematic variation with $\nu$, the
value of $\cl(\nu)/\bar{\cl} - 1$  is found to increases as we move from 
the nearest to the furthest end of the simulation box along LoS. This 
result correlates  well with the fact that $\xb$ increases along the 
LoS direction. This systematic variation, however, is only seen at the 
large $\ell$ bins (small angular scales) where we have a large number of 
Fourier modes  in each bin. This systematic variation is not seen in the 
small $\ell$ bins, partly because of the fewer number of Fourier modes in 
each bin (leading to large sample variance) and partly due to the fact that 
the evolution of the \HI signal at these scales is dominated by the evolution 
of ionised bubbles not the $\xb$. To avoid this uncertainty, we restrict 
the $\ell$ range to $\ell>2571$ for the present analysis.

\begin{figure}
\psfrag{ratio}[c][c][1][0]{\large $A\,\sqrt{\cl(\nu)/\bar{\cl}}$}
\psfrag{nu}[c][c][1][0]{\large $\nu - \nu_{\rm c}$\,~MHz}
\psfrag{diff}[c][c][1][0]{\large Residual}
\psfrag{fit}[c][c][1][0]{\large Fit}
\psfrag{reionization history}[c][c][1][0]{\large Reionization history~~~~}
\centering
\includegraphics[width=0.48\textwidth, angle=0]{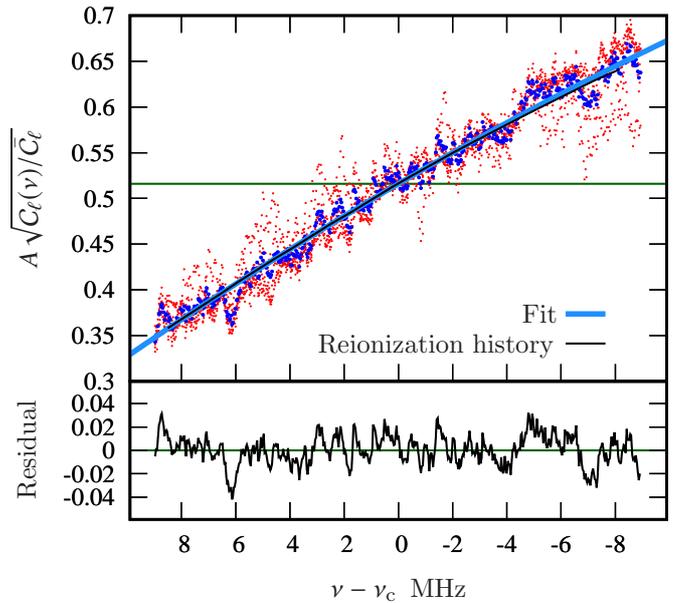}
\caption{The upper panel shows $A\,\sqrt{\cl(\nu)/\bar{\cl}}$ estimated 
from our LC EoR simulation. The red points show the results at different 
$\ell$ bins with $\ell>2571$. The blue points show  $A\,\sqrt{\c(\nu)/
\bar{\c}}$ which have been estimated by combining all the modes with 
$\ell>2571$ into a single bin. The light blue solid line shows the 
$2^{\rm nd}$ order polynomial fit to the values of $A \sqrt{\c(\nu)
/\bar{\c}}$ and the lower panel shows the residual after subtracting out 
the fit. The black solid line shows the values of $\xb(\nu)$ corresponding 
to the reionization history of our LC simulation. The horizontal green
line shows the value $\xb(\nu_{\rm c})=0.51$.}
\label{fig:cl_ratio_LC}
\end{figure}

We divide the $\ell$ range corresponding to our LC simulation into equally 
spaced logarithmic bins, and we compute $\sqrt{\cl(\nu) 
/\bar{\cl}}$ for all of these bins for which  $\ell>2571$. We see that 
the values of $\sqrt{\cl(\nu)/\bar{\cl}}$ display a large scatter 
(Fig.~\ref{fig:cl_ratio_LC}) {\it i.e.} for a fixed $\nu$ we find a 
range of values of $\sqrt{\cl(\nu)/\bar{\cl}}$ across the different 
$\ell$ bins. We attempt to mitigate the effect of these  variations 
by estimating $\c(\nu)$ which is obtained by combining the signal in 
all the modes with  $\ell>2571$ into a single bin. Note that the EoR 
21-cm signal is highly non-Gaussian \citep{bharadwaj05a, mondal15,
majumdar18} which makes it quite  non-trivial to predict errors for
the estimated $\c(\nu)$ \citep{mondal16}, and we have not attempted 
this here. Apart from the cosmic variance there will be instrumental
noise which will further worsen our predictions (i.e. the goodness
of fit $\chi^2$ of our model).  

We find that in addition to a systematic increase with decreasing frequency,
the values of $\sqrt{\c(\nu)/\bar{\c}}$ exhibit an apparently random 
fluctuation with varying frequency (Fig.~\ref{fig:cl_ratio_LC}). In order 
to model this systematic variation we have fitted a $2^{\rm nd}$ order 
polynomial $\sqrt{\c(\nu)/\bar{\c}}=a_0 + a_1 \left(\frac{\nu-\nu_{\rm c}}{B}
\right)+ a_2 \left(\frac{\nu-\nu_{\rm c}}{B}\right)^2 $ to the values 
estimated from the LC simulation. We have used a least-squares fit to 
obtain the best fit $a_0$, $a_1$ and $a_2$. We find that the best 
fit curve captures the systematic variation of $\sqrt{\c(\nu)/\bar{\c}}$ 
quite well and the residuals after subtracting out the fit appear to be
consistent with random fluctuations around zero (lower panel of
Fig.~\ref{fig:cl_ratio_LC}). If our model (eq.~\ref{eq:a4}) holds we
then have $\xb(\nu)=A \sqrt{\c(\nu)/\bar{\c}}$. As mentioned earlier,
it is necessary to introduce one additional input to determine the value
of $A$. Here we use the information that we have $\xb \approx 0.51$
at $\nu_{\rm c}=157.78 \,{\rm MHz}$. We use this in conjunction with
the polynomial fit to determine the value of $A$. Fig.~\ref{fig:cl_ratio_LC} 
shows a comparison of $\xb(\nu)$ corresponding to the reionization 
history (see Fig.~2 of \citealt{mondal18}) of our LC simulation and 
the best fit values of $A \sqrt{\c(\nu)/\bar{\c}}$ estimated from the 
LC simulation. We find that the two are in close agreement, thereby 
validating our model. 


\section{Summary and Conclusions}
The LC effect imprints the cosmological evolution history on the redshifted 
HI 21-cm signal $T_{\rm b} (\n, \nu)$ along the LoS direction $\nu$. This 
effect is particularly pronounced during EoR when $\xb(\nu)$ falls rapidly 
as the universe evolves. The MAPS $\cl(\nu_1, \nu_2)$ fully quantifies the 
second order statistics of $T_{\rm b} (\n, \nu)$. It does not assume the 
signal to be ergodic along the LoS direction $\nu$, and the frequency 
$(\nu_1,\nu_2)$ dependence of $\cl(\nu_1, \nu_2)$ quantifies both the 
systematic variation and the random fluctuations of the signal along $\nu$. 
Here we have proposed a simple model (eq.~\ref{eq:a4}) where the systematic 
variations of $\cl(\nu_1, \nu_2)$ with $(\nu_1, \nu_2)$ arise entirely 
due to the evolution of $\xb(\nu)$. This provides an unique method to 
observationally  determine the reionization history of the universe. 

In this Letter we have used a LC simulation of the EoR 21-cm signal 
to estimate $\cl(\nu_1,\nu_2)$. Using the diagonal elements $\cl(\nu) 
\equiv \cl(\nu,\nu)$ we show that our model (eq.~\ref{eq:a4}) is indeed 
valid for large values of $\ell$. Assuming an external input which provides
us with the value of $\xb(\nu_c)$ at a particular frequency $\nu_c$, we 
demonstrate that it is possible to recover the reionization history 
$\xb(\nu)$ from the estimated $\cl(\nu)$ across the entire observational 
bandwidth $B$. The accuracy of our estimates depends on how accurately 
the value of $\xb$ is measured at a particular frequency. An incorrect 
determination of $\xb$ will results in a biased estimate of the 
reionization history.

The present analysis of $\cl(\nu_1,\nu_2)$ is restricted to the diagonal 
elements $(\nu_1=\nu_2)$. The analysis can be enlarged to include the 
information contained in the non-diagonal elements and thereby improve 
the signal to noise ratio for the recovered $\xb(\nu)$. It is however 
necessary to note that the EoR 21-cm signal is largely localised in the 
elements within the vicinity of the diagonal elements, and the elements 
at a large frequency separation $\mid \nu_1-\nu_2 \mid$ do not contain 
significant signal \citep{bharadwaj05,datta07a}. We plan to address these 
issues in future work.

Our analysis is a proof of concept and based on simple semi-numerical 
simulations. The details will possibly differ if one uses high resolution 
simulations or includes fully coupled 3D radiative transfer (e.g. 
\citealt{iliev06a, gnedin16}). However, one can treat our predictions 
as being characteristic of the qualitative nature of the non-ergodic 
LC EoR 21-cm signal.


\section*{Acknowledgements}
This work was supported by the Science and Technology Facilities Council
[grant numbers ST/F002858/1 and ST/I000976/1] and the Southeast Physics
Network~(SEPNet).


\bibliographystyle{mnras} 
\bibliography{refs}



\label{lastpage}

\end{document}